\input harvmac
%

\let\includefigures=\iftrue
\let\useblackboard=\iftrue
\newfam\black

\includefigures
\message{If you do not have epsf.tex (to include figures),}
\message{change the option at the top of the tex file.}
\input epsf
\def\figin{\epsfcheck\figin}\def\figins{\epsfcheck\figins}
\def\epsfcheck{\ifx\epsfbox\UnDeFiNeD
\message{(NO epsf.tex, FIGURES WILL BE IGNORED)}
\gdef\figin##1{\vskip2in}\gdef\figins##1{\hskip.5in}
\else\message{(FIGURES WILL BE INCLUDED)}%
\gdef\figin##1{##1}\gdef\figins##1{##1}\fi}
\def\DefWarn#1{}
\def\figinsert{\goodbreak\midinsert}
\def\ifig#1#2#3{\DefWarn#1\xdef#1{fig.~\the\figno}
\writedef{#1\leftbracket fig.\noexpand~\the\figno}%
\figinsert\figin{\centerline{#3}}\medskip\centerline{\vbox{
\baselineskip12pt\advance\hsize by -1truein
\noindent\footnotefont{\bf Fig.~\the\figno:} #2}}
\endinsert\global\advance\figno by1}
\else
\def\ifig#1#2#3{\xdef#1{fig.~\the\figno}
\writedef{#1\leftbracket fig.\noexpand~\the\figno}%
\global\advance\figno by1} \fi

\def\journal#1&#2(#3){\unskip, \sl #1\ \bf #2 \rm(19#3) }
\def\andjournal#1&#2(#3){\sl #1~\bf #2 \rm (19#3) }

\def\ie{{\it i.e.}}
\def\eg{{\it e.g.}}

\noblackbox
%


\def\unlockat{\catcode`\@=11}
\def\lockat{\catcode`\@=12}

\unlockat

\def\newsec#1{\global\advance\secno by1\message{(\the\secno. #1)}
\global\subsecno=0\global\subsubsecno=0\eqnres@t\noindent
{\bf\the\secno. #1}
\writetoca{{\secsym} {#1}}\par\nobreak\medskip\nobreak}
\global\newcount\subsecno \global\subsecno=0
\def\subsec#1{\global\advance\subsecno
by1\message{(\secsym\the\subsecno. #1)}
\ifnum\lastpenalty>9000\else\bigbreak\fi\global\subsubsecno=0
\noindent{\it\secsym\the\subsecno. #1}
\writetoca{\string\quad {\secsym\the\subsecno.} {#1}}
\par\nobreak\medskip\nobreak}
\global\newcount\subsubsecno \global\subsubsecno=0
\def\subsubsec#1{\global\advance\subsubsecno by1
\message{(\secsym\the\subsecno.\the\subsubsecno. #1)}
\ifnum\lastpenalty>9000\else\bigbreak\fi
\noindent\quad{\secsym\the\subsecno.\the\subsubsecno.}{#1}
\writetoca{\string\qquad{\secsym\the\subsecno.\the\subsubsecno.}{#1}}
\par\nobreak\medskip\nobreak}

\def\subsubseclab#1{\DefWarn#1\xdef
#1{\noexpand\hyperref{}{subsubsection}%
{\secsym\the\subsecno.\the\subsubsecno}%
{\secsym\the\subsecno.\the\subsubsecno}}%
\writedef{#1\leftbracket#1}\wrlabeL{#1=#1}}
\lockat

\def\ie{{\it i.e.}}
\def\eg{{\it e.g.}}


\font\manual=manfnt \def\dbend{\lower3.5pt\hbox{\manual\char127}}

\def\IZ{\relax\ifmmode\mathchoice
{\hbox{\cmss Z\kern-.4em Z}}{\hbox{\cmss Z\kern-.4em Z}}
{\lower.9pt\hbox{\cmsss Z\kern-.4em Z}}
{\lower1.2pt\hbox{\cmsss Z\kern-.4em Z}}\else{\cmss Z\kern-.4em
Z}\fi}
\def\half{{1\over 2}}


\def\IZ{\relax\ifmmode\mathchoice
{\hbox{\cmss Z\kern-.4em Z}}{\hbox{\cmss Z\kern-.4em Z}}
{\lower.9pt\hbox{\cmsss Z\kern-.4em Z}}
{\lower1.2pt\hbox{\cmsss Z\kern-.4em Z}}\else{\cmss Z\kern-.4em
Z}\fi}
\def\IB{\relax{\rm I\kern-.18em B}}
\def\IC{{\relax\hbox{$\inbar\kern-.3em{\rm C}$}}}
\def\ID{\relax{\rm I\kern-.18em D}}
\def\IE{\relax{\rm I\kern-.18em E}}
\def\IF{\relax{\rm I\kern-.18em F}}
\def\IG{\relax\hbox{$\inbar\kern-.3em{\rm G}$}}
\def\IGa{\relax\hbox{${\rm I}\kern-.18em\Gamma$}}
\def\IH{\relax{\rm I\kern-.18em H}}
\def\II{\relax{\rm I\kern-.18em I}}
\def\IK{\relax{\rm I\kern-.18em K}}
\def\IP{\relax{\rm I\kern-.18em P}}
\def\IQ{\relax\hbox{$\inbar\kern-.3em{\rm Q}$}}

\def\inbar{\,\vrule height1.5ex width.4pt depth0pt}

\font\cmss=cmss10 \font\cmsss=cmss10 at 7pt
\def\IR{\relax{\rm I\kern-.18em R}}

%
%

\def\makeblankbox#1#2{\hbox{\lower\dp0\vbox{\hidehrule{#1}{#2}%
   \kern -#1
   \hbox to \wd0{\hidevrule{#1}{#2}%
      \raise\ht0\vbox to #1{}
      \lower\dp0\vtop to #1{}
      \hfil\hidevrule{#2}{#1}}%
   \kern-#1\hidehrule{#2}{#1}}}%
}%
\def\hidehrule#1#2{\kern-#1\hrule height#1 depth#2 \kern-#2}%
\def\hidevrule#1#2{\kern-#1{\dimen0=#1\advance\dimen0 by #2\vrule
    width\dimen0}\kern-#2}%
\def\openbox{\ht0=1.2mm \dp0=1.2mm \wd0=2.4mm  \raise 2.75pt
\makeblankbox {.25pt} {.25pt}  }

\def\bun#1/#2{\leavevmode
   \kern.1em \raise .5ex \hbox{\the\scriptfont0 #1}%
   \kern-.1em $/$%
   \kern-.15em \lower .25ex \hbox{\the\scriptfont0 #2}%
}

\def\opensquare{\ht0=3.4mm \dp0=3.4mm \wd0=6.8mm  \raise 2.7pt
\makeblankbox {.25pt} {.25pt}  }


\def\sector#1#2{\ {\scriptstyle #1}\hskip 1mm
\mathop{\opensquare}\limits_{\lower 1mm\hbox{$\scriptstyle#2$}}\hskip 1mm}

\def\tsector#1#2{\ {\scriptstyle #1}\hskip 1mm
\mathop{\opensquare}\limits_{\lower 1mm\hbox{$\scriptstyle#2$}}^\sim\hskip 1mm}


\def\inbar{\,\vrule height1.5ex width.4pt depth0pt}

\font\cmss=cmss10 \font\cmsss=cmss10 at 7pt
\def\IR{\relax{\rm I\kern-.18em R}}


\def\frac#1#2{{#1\over#2}}

\def\half{\frac12}

\def\inbar{\,\vrule height1.5ex width.4pt depth0pt}
\def\IC{\relax\hbox{$\inbar\kern-.3em{\rm C}$}}
\def\IR{\relax{\rm I\kern-.18em R}}
\def\IP{\relax{\rm I\kern-.18em P}}

%
%
\catcode`\@=11
\def\slash#1{\mathord{\mathpalette\c@ncel{#1}}}
\overfullrule=0pt

\def\II{{\cal I}}

\def\underrel#1\over#2{\mathrel{\mathop{\kern\z@#1}\limits_{#2}}}

\catcode`\@=12


%

\def\cosh{{\rm cosh}}

\def\exp{{\rm exp}}



\def\frac#1#2{{#1\over#2}}

\def\half{\frac12}

\def\inbar{\,\vrule height1.5ex width.4pt depth0pt}
\def\IC{\relax\hbox{$\inbar\kern-.3em{\rm C}$}}
\def\IR{\relax{\rm I\kern-.18em R}}
\def\IP{\relax{\rm I\kern-.18em P}}

%
%

%
\catcode`\@=11
\def\slash#1{\mathord{\mathpalette\c@ncel{#1}}}
\overfullrule=0pt

\def\II{{\cal I}}

\def\underrel#1\over#2{\mathrel{\mathop{\kern\z@#1}\limits_{#2}}}

\catcode`\@=12


%

\def \cosh{{\rm cosh}}

\def\exp{{\rm exp}}



\lref\BekensteinUR{
  J.~D.~Bekenstein,
  ``Black holes and entropy,''
Phys.\ Rev.\ D {\bf 7}, 2333 (1973).
}

\lref\MarolfDBA{
  D.~Marolf and J.~Polchinski,
  ``Gauge/Gravity Duality and the Black Hole Interior,''
Phys.\ Rev.\ Lett.\  {\bf 111}, 171301 (2013).
[arXiv:1307.4706 [hep-th]].
}

\lref\PapadodimasJKU{
  K.~Papadodimas and S.~Raju,
  ``State-Dependent Bulk-Boundary Maps and Black Hole Complementarity,''
Phys.\ Rev.\ D {\bf 89}, 086010 (2014).
[arXiv:1310.6335 [hep-th]].
}

\lref\GiveonHFA{
  A.~Giveon, N.~Itzhaki and J.~Troost,
  ``Lessons on Black Holes from the Elliptic Genus,''
JHEP {\bf 1404}, 160 (2014).
[arXiv:1401.3104 [hep-th]].
}

\lref\HawkingSW{
  S.~W.~Hawking,
  ``Particle Creation by Black Holes,''
Commun.\ Math.\ Phys.\  {\bf 43}, 199 (1975), [Erratum-ibid.\  {\bf 46}, 206 (1976)].
}

\lref\AtickSI{
  J.~J.~Atick and E.~Witten,
  ``The Hagedorn Transition and the Number of Degrees of Freedom of String Theory,''
Nucl.\ Phys.\ B {\bf 310}, 291 (1988).
}

\lref\fzz{
V.A. Fateev, A.B.
Zamolodchikov and Al.B. Zamolodchikov, unpublished.
}

\lref\KarczmarekBW{
  J.~L.~Karczmarek, J.~M.~Maldacena and A.~Strominger,
  ``Black hole non-formation in the matrix model,''
JHEP {\bf 0601}, 039 (2006).
[hep-th/0411174].
}

\lref\AharonyXN{
  O.~Aharony, A.~Giveon and D.~Kutasov,
  ``LSZ in LST,''
Nucl.\ Phys.\ B {\bf 691}, 3 (2004).
[hep-th/0404016].
}

\lref\TeschnerUG{
  J.~Teschner,
  ``Operator product expansion and factorization in the H+(3) WZNW model,''
Nucl.\ Phys.\ B {\bf 571}, 555 (2000).
[hep-th/9906215].
}

\lref\MaldacenaRE{
  J.~M.~Maldacena,
  ``The Large N limit of superconformal field theories and supergravity,''
Int.\ J.\ Theor.\ Phys.\  {\bf 38}, 1113 (1999), [Adv.\ Theor.\ Math.\ Phys.\  {\bf 2}, 231 (1998)].
[hep-th/9711200].
}

\lref\tHooftGX{
  G.~'t Hooft,
  ``Dimensional reduction in quantum gravity,''
Salamfest 1993:0284-296.
[gr-qc/9310026].
}

\lref\HawkingRA{
  S.~W.~Hawking,
  ``Breakdown of Predictability in Gravitational Collapse,''
Phys.\ Rev.\ D {\bf 14}, 2460 (1976).
}

\lref\SusskindVU{
  L.~Susskind,
  ``The World as a hologram,''
J.\ Math.\ Phys.\  {\bf 36}, 6377 (1995).
[hep-th/9409089].
}

\lref\GiveonPX{
  A.~Giveon and D.~Kutasov,
  ``Little string theory in a double scaling limit,''
JHEP {\bf 9910}, 034 (1999).
[hep-th/9909110].
}

\lref\DijkgraafBA{
  R.~Dijkgraaf, H.~L.~Verlinde and E.~P.~Verlinde,
  ``String propagation in a black hole geometry,''
Nucl.\ Phys.\ B {\bf 371}, 269 (1992).
}

\lref\KazakovPM{
  V.~Kazakov, I.~K.~Kostov and D.~Kutasov,
  ``A Matrix model for the two-dimensional black hole,''
Nucl.\ Phys.\ B {\bf 622}, 141 (2002).
[hep-th/0101011].
}

\lref\KutasovRR{
  D.~Kutasov,
  ``Accelerating branes and the string/black hole transition,''
[hep-th/0509170].
}

\lref\ElitzurCB{
  S.~Elitzur, A.~Forge and E.~Rabinovici,
  ``Some global aspects of string compactifications,''
Nucl.\ Phys.\ B {\bf 359}, 581 (1991).
}

\lref\WittenYR{
  E.~Witten,
  ``On string theory and black holes,''
Phys.\ Rev.\ D {\bf 44}, 314 (1991).
}

\lref\KazamaQP{
  Y.~Kazama and H.~Suzuki,
 ``New N=2 Superconformal Field Theories and Superstring Compactification,''
Nucl.\ Phys.\ B {\bf 321}, 232 (1989).
}

\lref\MandalTZ{
  G.~Mandal, A.~M.~Sengupta and S.~R.~Wadia,
  ``Classical solutions of two-dimensional string theory,''
Mod.\ Phys.\ Lett.\ A {\bf 6}, 1685 (1991).
}

\lref\GiveonKP{
  A.~Giveon and N.~Itzhaki,
  ``String Theory Versus Black Hole Complementarity,''
JHEP {\bf 1212}, 094 (2012).
[arXiv:1208.3930 [hep-th]].
}

\lref\GiveonICA{
  A.~Giveon and N.~Itzhaki,
  ``String theory at the tip of the cigar,''
JHEP {\bf 1309}, 079 (2013).
[arXiv:1305.4799 [hep-th]].
}

\lref\GiveonHSA{
  A.~Giveon, N.~Itzhaki and J.~Troost,
  ``The Black Hole Interior and a Curious Sum Rule,''
JHEP {\bf 1403}, 063 (2014).
[arXiv:1311.5189 [hep-th]].
}

\lref\MertensPZA{
  T.~G.~Mertens, H.~Verschelde and V.~I.~Zakharov,
  ``Near-Hagedorn Thermodynamics and Random Walks: a General Formalism in Curved Backgrounds,''
JHEP {\bf 1402}, 127 (2014).
[arXiv:1305.7443 [hep-th]].
}

\lref\MertensZYA{
  T.~G.~Mertens, H.~Verschelde and V.~I.~Zakharov,
  ``Random Walks in Rindler Spacetime and String Theory at the Tip of the Cigar,''
JHEP {\bf 1403}, 086 (2014).
[arXiv:1307.3491 [hep-th]].
}

\lref\MertensDIA{
  T.~G.~Mertens, H.~Verschelde and V.~I.~Zakharov,
  ``On the Relevance of the Thermal Scalar,''
JHEP {\bf 1411}, 157 (2014).
[arXiv:1408.7012 [hep-th]].
}

\lref\MertensSAA{
  T.~G.~Mertens, H.~Verschelde and V.~I.~Zakharov,
  ``Perturbative String Thermodynamics near Black Hole Horizons,''
[arXiv:1410.8009 [hep-th]].
}

\lref\GiveonMI{
  A.~Giveon, D.~Kutasov, E.~Rabinovici and A.~Sever,
  ``Phases of quantum gravity in AdS(3) and linear dilaton backgrounds,''
Nucl.\ Phys.\ B {\bf 719}, 3 (2005).
[hep-th/0503121].
}

\lref\AharonyVK{
  O.~Aharony, B.~Fiol, D.~Kutasov and D.~A.~Sahakyan,
  ``Little string theory and heterotic / type II duality,''
Nucl.\ Phys.\ B {\bf 679}, 3 (2004).
[hep-th/0310197].
}

\lref\AharonyUB{
  O.~Aharony, M.~Berkooz, D.~Kutasov and N.~Seiberg,
  ``Linear dilatons, NS five-branes and holography,''
JHEP {\bf 9810}, 004 (1998).
[hep-th/9808149].
}

\lref\SeibergBJ{
  N.~Seiberg and S.~H.~Shenker,
  ``A Note on background (in)dependence,''
Phys.\ Rev.\ D {\bf 45}, 4581 (1992).
[hep-th/9201017].
}

\Title{
} {\vbox{
\bigskip\centerline{Stringy Horizons}}}
\medskip
\centerline{\it Amit Giveon${}^{1}$, Nissan Itzhaki${}^{2}$ and David Kutasov${}^{3}$}
\bigskip
\smallskip
\centerline{${}^{1}$Racah Institute of Physics, The Hebrew
University} \centerline{Jerusalem 91904, Israel}
\smallskip
\centerline{${}^{2}$ Physics Department, Tel-Aviv University, Israel} \centerline{Ramat-Aviv, 69978, Israel}
\smallskip
\centerline{${}^3$EFI and Department of Physics, University of
Chicago} \centerline{5640 S. Ellis Av., Chicago, IL 60637, USA }

\bigskip\bigskip\bigskip
\noindent

We argue that classical $(\alpha')$ effects qualitatively modify the structure of Euclidean black hole horizons in string theory. While low energy modes experience the geometry familiar from general relativity, high energy ones see a rather different geometry, in which the Euclidean horizon can be penetrated by an amount that grows with the radial momentum of the probe. We discuss this in the exactly solvable $SL(2,R)/U(1)$ black hole, where it is a manifestation of the black hole/Sine-Liouville duality.

\vglue .3cm
\bigskip

\Date{2/15}

\bigskip

\newsec{Introduction}

About forty years ago, J. Bekenstein \BekensteinUR\ and S. Hawking \HawkingSW\ proposed that in semiclassical quantum gravity black holes behave like thermodynamic objects. In particular, they appear to have an entropy that is proportional to their area and satisfies the usual thermodynamic laws. Thus, black holes behave quantum mechanically like black bodies with  finite temperature.

It was recognized early on that the above successes came with a heavy price tag. The fact that black holes seem to behave as thermal objects suggests that quantum unitarity may be lost in their vicinity. In particular, Hawking showed \HawkingRA\ that one can think of the  finite temperature of a black hole as due to a process occurring near the horizon, where a pair of particles is created, one member of the pair falls into the horizon while the other escapes to infinity as thermal radiation. Thus, information associated with a pure state of a quantum system in the vicinity of a black hole horizon can be destroyed when the system falls into the black hole, which then evaporates via emission of Hawking radiation.

It is expected that corrections to Hawking's one-loop calculation should restore unitarity, with the information associated with the initial quantum state encoded in the outgoing radiation, but it is not clear how this works. In Hawking's calculation, the horizon plays a crucial role, but in classical general relativity it is not a special place and, in particular, an infalling observer is not expected to see anything unusual while crossing it. Thus, it is unclear how quantum unitarity is restored.

It was initially hoped that gauge/gravity duality \MaldacenaRE\ will lead to a resolution of these problems, however, so far no universally accepted picture emerged, despite much work on the subject (see \eg\ \refs{\MarolfDBA,\PapadodimasJKU} for some recent discussions). Part of the problem appears to be that gauge/gravity duality maps gravity to the thermodynamics of the field theory, and in order to understand in detail why/how unitarity is preserved one has to go beyond the thermodynamic description. In the gauge theory, one must include in the analysis the degrees of freedom that describe generic high energy states. In the dual bulk theory, one must go beyond the classical gravity approximation. In all known examples, the underlying bulk theory is a string theory (which may be strongly coupled). Thus, one is led to study string theory in the background of black holes, and search for qualitative new effects associated with black hole horizons.

In weakly coupled string theory, one can imagine two types of such effects: quantum ($g_s$) effects, which are sensitive to the Planck length $l_p$, and string ($\alpha'$) effects, that are sensitive to the string length $l_s=\sqrt{\alpha'}$. In fact, it often happens in string theory that effects that from the gravity perspective should appear at $l_p$ actually appear at $l_s$. A famous example is the value of the effective UV cutoff in perturbative string loop amplitudes.  Thus, it is natural to ask whether {\it classical } string effects play a role in resolving the puzzles of black hole physics. This question motivated the present study.

Since we are interested in phenomena that occur near the horizon, it is natural to expect that they should be visible in the Euclidean Black Hole (EBH) geometry, which has the advantage of being free of curvature singularities. The non-trivial part of any EBH spacetime is the two-dimensional geometry in the Euclidean time and radial directions. At infinity in the radial direction it typically approaches a cylinder, whose circumference (in the Euclidean time direction) is the inverse temperature. As one moves towards the origin, the radius of this circle decreases and it eventually closes up into a smooth semi-infinite cigar. The tip of the cigar corresponds (upon Wick rotation) to the black hole horizon.

In this geometry, it is natural to consider a closed string which winds once around the circle at large $r$. The lowest lying state of such a string is the tachyon, which turns out to be in the spectrum, even in theories with a chiral GSO projection, in which the tachyon with winding number zero is projected out. The presence of the winding tachyon in finite temperature string theory was noted early on in the development of the subject; see \eg\ \AtickSI, where it was suggested that it might be related to a phase transition that occurs at or near the Hagedorn temperature.

More recently, it was argued in \KutasovRR\ that in string theory in a EBH background, this wound tachyon has a non-zero expectation value. This agrees with the phase transition picture, as the black hole describes the phase in which the winding symmetry around Euclidean time is broken (as is obvious from the fact that the Euclidean time circle is contractible). The expectation value of the wound tachyon can be thought of as a non-perturbative $\alpha'$ effect in classical string theory, and it is natural to ask how it modifies the properties of the (Euclidean) horizon from GR expectations. This, and related questions, were further discussed in the last couple of years in \refs{\GiveonKP\GiveonICA\MertensPZA\MertensZYA\GiveonHSA\GiveonHFA\MertensDIA-\MertensSAA}, but remain largely open. The main goal of this note\foot{And additional work that will appear separately.}  is to shed additional light on the subject.

To do this, we will view the EBH geometry as a spatial background, and add to it a time direction. We will then consider scattering states on the asymptotic cylinder, and study the phase shift for scattering off the tip of the cigar. The idea is to use this to probe the effects of the tachyon condensate, which is localized near the tip.

The discussion above applies to any black holes in string theory, but in general it is hard to complete this program, since the worldsheet sigma model for the black hole is not sufficiently well understood. Thus, we will specialize to a specific type of black hole, the two-dimensional black hole corresponding to the $SL(2,R)/U(1)$ coset, discovered in \refs{\ElitzurCB\MandalTZ-\WittenYR}. This black hole lives in a two-dimensional spacetime with asymptotic spatial linear dilaton, and plays a number of important roles in fivebrane physics and other contexts (see \eg\ \refs{\AharonyVK,\AharonyXN} for  reviews). Its important features for our purposes are:
\item{(1)} The classical reflection coefficient for scattering states is known exactly in $\alpha'$ in this case \TeschnerUG.
\item{(2)} The fact that in addition to the cigar geometry the worldsheet CFT involves a condensate of the wound tachyon has been established in this case; it is known as FZZ duality \refs{\fzz,\KazakovPM}.

\noindent
In the rest of this note, we will study the above reflection coefficient and, in particular, the string corrections to the gravity result. We will then attempt to use it to learn about the near-horizon structure of this black hole.

\newsec{The $SL(2,R)/U(1)$ black hole}

The $SL(2,R)/U(1)$ coset CFT describes string propagation on a semi-infinite cigar, with metric and dilaton\foot{We discuss the coset that gives the Euclidean black hole. One can also consider the Minkowski black hole, but we will not do that here.}
\eqn\sltwo{\eqalign{&ds^2=k(dr^2+\tanh^2rd\theta^2)~;\cr
&\Phi-\Phi_0=-\ln\cosh r~.}}
$\theta\sim \theta+2\pi$ is an angular coordinate, while $0\le r<\infty$ is the direction along the cigar; $r=0$ is the tip. The string coupling $e^\Phi$ depends on $r$; it goes to zero far from the tip and attains its maximal value, $e^{\Phi_0}$, at the tip.   $k$ is a free parameter, which governs the overall size of the cigar. In the algebraic coset description, it corresponds to the level of the underlying $SL(2,R)$ current algebra. Geometrically, it sets the overall scale of the cigar. In particular, at large $k$ \sltwo\ describes a large, weakly curved geometry. That's the analog in this context of a large (Euclidean) Schwarzschild black hole in higher dimensions.

The model comes in two versions, depending on whether one is studying it in the bosonic string or the superstring. In the former case, one is interested in the bosonic coset model, whose central charge is given by
\eqn\bosc{c=2+{6\over k-2}~.}
The background fields \sltwo\ are expected in this case to receive perturbative $\alpha'$ corrections, which can be thought of as $1/k$ corrections. These corrections are not expected to play an important role in our story.

In the superstring, one is interested in the $N=1$ superconformal coset,\foot{Which happens to have $N=2$ superconformal symmetry; this is an example of the Kazama-Suzuki \KazamaQP\ construction.} which is obtained by attaching to a bosonic $SL(2,R)$ WZW model three free fermions that transform in the adjoint representation of $SL(2,R)$, and gauging a diagonal $U(1)$ in the full $SL(2,R)$ of bosons $+$ fermions. The total level of $SL(2,R)$, $k$, can be written in this case as a sum of bosonic and fermionic contributions, $k=(k+2)+(-2)$, and the corresponding central charge is
\eqn\supc{c=3+{6\over k}~.}
In this case, the background \sltwo\ is not expected to receive perturbative corrections in $1/k$.

Although, as usual in string theory, to talk about a well defined theory with a stable vacuum one needs to consider the superstring, for our purposes the bosonic theory is good enough, since the physics we are interested in is unrelated to the usual closed string tachyon. Hence, we will phrase the discussion below in this language; it is straightforward to repeat it in the worldsheet supersymmetric case.

Momentum and winding modes on the cigar correspond to the operators $V_{j;m,\bar m}$, whose scaling dimensions are given by
\eqn\delbar{\eqalign{
\Delta_{j;m,\bar m}&=-{j(j+1)\over k-2}+{m^2\over k}~;\cr
\bar\Delta_{j;m,\bar m}&=-{j(j+1)\over k-2}+{\bar m^2\over k}~.
}}
The quantum number $j$ can be real or complex. The former plays an important role in describing normalizable states on the cigar and off-shell observables important for linear dilaton holography \AharonyUB. The latter is useful for describing scattering states on the cigar, for which one has
\eqn\scatstat{j=-\half+i\hat s~.}
Looking back at the dimension formula \delbar, we see that the radial momentum conjugate to the well normalized coordinate $\phi=\sqrt{k}r$ is
\eqn\pphhii{p_\phi=2\hat s\sqrt{1\over k-2}~.}
$m$ and $\bar m$ label momentum and winding around the cigar; they take the values
\eqn\momwin{\eqalign{m=&\half(wk+p)~,\cr
                                  \bar m=&\half(wk-p)~.
                                  }}
The integers $p$ and $w$ are the momentum and winding around the circle.  Eq. \delbar\ is compatible with the radius of the circle being  $R^2=\alpha' k$, in agreement with  \sltwo\ (which is written for $\alpha'=1$, a convention that we will continue to use below).

An important observation for our purposes is that the model described above is related to one that superficially is quite different, the Sine-Liouville theory described by the Lagrangian
\eqn\lsinel{\CL={1\over 4\pi}\left[(\partial x)^2+(\partial\phi)^2+Q\hat R\phi+\lambda e^{b\phi}\cos R(x_L-x_R)\right]~.
}
This Lagrangian describes string propagation on an infinite cylinder of radius $R$, with a potential that goes to zero as $\phi\to\infty$ (since $b$ is negative) and blows up as $\phi\to\infty$. The dilaton is linear in $\phi$; the string coupling goes to zero as $\phi\to\infty$ and diverges as $\phi\to-\infty$. However, the potential prevents the string from exploring this region (see below for further discussion of this issue).

In the weakly coupled region $\phi\to\infty$,  the backgrounds \sltwo, \lsinel\ agree, if we take $Q=1/\sqrt{k-2}$, $R=\sqrt k$, $b=-1/Q$. However, at finite $\phi$ they superficially look different. The former has a hard wall at $\phi=0$, the end of space, while the latter has a soft wall in the sense that the higher the energy of a string, the farther it can penetrate towards the strong coupling region. This is similar to what happens in Liouville theory; see \eg\ \SeibergBJ\ for a discussion.

Nevertheless, it has been argued  that these two models are in fact equivalent. In the bosonic case, discussed here, this was proposed based on worldsheet considerations in unpublished work by V. Fateev, A. and Al. Zamolodchikov \fzz; see \KazakovPM\ for a review.  In the supersymmetric case, where instead of Sine-Liouville one has $N=2$ Liouville, it was  proposed in \GiveonPX, who arrived at it from a spacetime perspective, by studying fivebrane physics.

This equivalence is usually viewed as a weak-strong coupling duality on the worldsheet. At large $k$, the cigar sigma model \sltwo\ is weakly coupled (in the sense of the $\alpha'$ expansion), while the Sine-Liouville model  \lsinel\ is strongly coupled, since the slope of the linear dilaton, $Q$, is small, and the Sine-Liouville potential is steep. On the other hand, for large $Q$, the cigar is strongly coupled, while the Sine-Liouville is weakly coupled. For general $Q$ of order one, one has to take into account both the cigar geometry and the Sine-Liouville potential.

One can view the Sine-Liouville potential as an expectation value of the closed string tachyon with winding one around the $x$ circle. This operator is normalizable for sufficiently small $Q$ (or large $k$); as $Q$ increases ($k$ decreases), it eventually becomes non-normalizable \KarczmarekBW, leading to a phase transition discussed in \GiveonMI.

In this note,
we will discuss the large $k$ region. In that case, the conventional picture is that the physics is well described by the sigma model on the cigar \sltwo. The question we would like to address is what is the effect of the winding tachyon condensate \lsinel. We will do this in the next section, by studying the phase shift for scattering of the states \scatstat\ from the tip of the cigar.

\newsec{Scattering on the cigar}

One of the main reasons we are studying the $SL(2,R)/U(1)$ black hole here is that in this case the phase shift for scattering is known exactly (to leading order in $g_s$). It is given by \refs{\TeschnerUG,\GiveonPX}
\eqn\phaseshift{e^{-i\delta}=\nu^{2j+1}{\Gamma(1-{2j+1\over k-2})\over\Gamma(1+{2j+1\over k-2})}
{\Gamma(-2j)\Gamma(j-m+1)\Gamma(1+j+\bar m)\over\Gamma(2j+2)\Gamma(-j-m)\Gamma(\bar m-j)}~.
}
Here $\nu$ is a real constant, which may depend on $k$, but not on the quantum numbers $(j;m,\bar m)$. The r.h.s. of \phaseshift\ is a phase, as can be seen by plugging in the values \scatstat,  \momwin\ for $(j;m,\bar m)$.

To understand the physics of \phaseshift, one can proceed as follows. For large $k$ and $p_\phi \ll \sqrt{k}$,
one can omit the first ratio of $\Gamma$ functions in \phaseshift. In this limit, one must reproduce the results of scattering in the geometry \sltwo\ in the gravity approximation
\DijkgraafBA; this is indeed the case.

It is instructive to take the limit of momentum large compared to the curvature of the background,
$p_{\phi} \gg Q$, in this approximation. One expects to find that the phase shift goes to zero in this limit, for the following reason. The curved background \sltwo\ gives rise to an effective potential for the incoming particle. However, as the radial momentum of the particle increases, this potential becomes less important, and the main effect is due to the fact that the radial direction is a half line $\phi\ge 0$. Thus, the phase shift approaches that of a free particle, which is by definition zero.

Alternatively, one can note that at large radial momentum in the gravity approximation, the phase shift receives contributions mostly from the vicinity of the tip of the cigar, which looks locally like $R^2$. The momentum around the circle $p$ \momwin\ can be thought of as angular momentum on $R^2$. Thus, the phase of the wavefunction for highly energetic particles should be the same as that for free particles on $R^2$, and since the phase shift is defined relative to that of a free particle, it must vanish in this limit.

Looking back at \phaseshift, the ratio of $\Gamma$  functions behaves in this limit as
\eqn\limgamma{{\Gamma(-2j)\Gamma(j-m+1)\Gamma(1+j+\bar m)\over\Gamma(2j+2)\Gamma(-j-m)\Gamma(\bar m-j)}\sim (-)^{\bar{m}-m}
e^{-2 i \hat s \ln4}~.
  }
The $(-)^{\bar{m}-m}=e^{-i\pi p}$ is the standard  angular momentum dependent phase for a free particle on $R^2$.  Thus, it does not contribute to the phase shift.
The last term in \limgamma\ is a contribution to the phase shift that is proportional to the radial momentum $p_\phi$, and we expect it to be absent for reasons mentioned in the previous paragraph.  Fortunately, the factor associated with $\nu$ in the phase shift  \phaseshift\ is proportional to $p_\phi$ as well, and we can tune $\nu$ such that the full phase shift $\delta$ goes to zero as $p_\phi\to\infty$ (in the gravity approximation).

So far, we ignored the contribution to the phase shift of the first ratio of $\Gamma$ functions in \phaseshift. As the radial momentum $p_\phi$ increases, this factor becomes more important. Eventually, for $\hat s \gg k$ (or $p_\phi\gg\sqrt k$), it comes to dominate the phase shift. In fact, using the Stirling formula one finds that the phase shift behaves in this limit like
\eqn\behavphase{\delta\simeq {4\hat s\over k-2} \ln(\hat s)\simeq 2 Q p_{\phi} \ln p_{\phi}~,}
where `$\simeq$' stands for an equality up to sub-leading terms at large $\hat s$, \ie\ $\delta$ goes like $p_\phi\ln p_\phi$ for large $p_\phi$. This is a rather surprising behavior. We argued before that the fact that the phase shift goes to zero at large radial momentum is due to very general features of the background \sltwo: the fact that the radial coordinate lives on the half line $\phi\ge 0$, and that the vicinity of the tip, $\phi=0$, the region that is presumably probed by highly energetic particles, looks like $R^2$.
From this point of view, the fact that the full phase shift \phaseshift\ does not go to zero, but instead grows at large $p_\phi$, suggests that high energy particles effectively see a different geometry than the cigar \sltwo, that potentially includes $\phi < 0$. Our goal is to try to understand that better.

The momenta for which this behavior is obtained are stringy, so it is not a priori obvious that a description in terms of particles scattering in a particular geometry is applicable. Therefore, in order to understand the behavior \behavphase, we need to go back to the worldsheet path integral and try to see
what is the origin of this behavior.

The following observation is useful to that end. As explained above, the large phase shift \behavphase\ comes from the first ratio of $\Gamma$ functions in \phaseshift. This ratio has poles for $(2j+1)/(k-2)=n$ with $n=1,2,\cdots$. As reviewed in \KazakovPM, these poles can be understood as coming from the Sine-Liouville interaction \lsinel. To be precise, the phase shift \phaseshift\ is related to the two point function $\langle V_{j;m,\bar m} V_{j;-m,-\bar m}\rangle$. In the Sine-Liouville model \lsinel, these operators correspond to
\eqn\equiv{V_{j;m,\bar m}\leftrightarrow e^{ip_L x_L+ip_Rx_R+\beta\phi}~,}
where $p_{L,R}$ are left and right-moving momenta on the circle, $p_L={p\over R}+wR$, $p_R={p\over R}-wR$, and $\beta=2Qj$.

To calculate the two  point function of the operators \equiv, one has to perform the path integral over $(\phi,x)$ with the action \lsinel. The path integral over the Liouville field $\phi$ can be split into an ordinary integral over the zero mode $\phi_0$, and the path integral over the non-zero modes $\phi_1(z,\bar z)$, with $\phi(z,\bar z)=\phi_0+\phi_1(z,\bar z)$. The zero mode integral takes the form
\eqn\intzero{
\int_{-\infty}^\infty d\phi_0 \exp\left[2\beta\phi_0+2Q\phi_0-\lambda e^{b\phi_0}\int e^{b\phi_1}\cos R(x_L-x_R)\right].
}
The first term in the exponential  comes from the two insertions of vertex operators, the second from the curvature coupling on the sphere, and the third from the Sine-Liouville interaction. The integral over $\phi_0$ can be performed exactly \KazakovPM, and gives (up to an unimportant overall constant)
\eqn\xxxint{\left(\lambda\int e^{b\phi_1}\cos R(x_L-x_R)\right)^s\Gamma(-s)~,}
where
\eqn\ssss{s=-2(\beta+Q)/b={2(2j+1)\over k-2}~.}
The poles at $2j+1=n(k-2)$, discussed above, originate from the poles of the factor $\Gamma(-s)$ in \xxxint. They are bulk poles in the sense of \AharonyXN, \ie\ their residues receive contributions from the bulk of $\phi_0$ space, far from the Sine-Liouville wall,  and can be computed by evaluating the free path integral in that region; see \KazakovPM\ for a discussion.

Coming back to the large phase shift \behavphase, since its origin is the same ratio of $\Gamma$ functions, it is natural to expect that it too is due to the Sine-Liouville interaction. We next argue that making this assumption gives the correct answer.

Plugging \scatstat\ into \ssss, we see that in this case the parameter $s$ is imaginary,
\eqn\sshat{s={4i\hat s\over k-2}~.}
The zero mode integral \intzero\ gives in this case \xxxint\ with this value of $s$. The non-zero mode path integral (over $\phi_1$) involves the action \lsinel\ with $\lambda=0$, the insertion of the two vertex operators \equiv, and the prefactor  $\lambda^s\Gamma(-s)$ in \xxxint.

For large $s$, we can use the Stirling approximation to evaluate the zero mode contribution
\eqn\sa{\Gamma(-s) \sim e^{-{4i\hat s\over k-2} \log(\hat s)} e^{-{2 \pi \hat s\over k-2}}= e^{-2i Q p_{\phi} \log(p_{\phi}) } e^{-{Q\pi p_{\phi}}}~.}
Interestingly, the first exponent is exactly the same phase as obtained from the first ratio of $\Gamma$ functions in \phaseshift, which can be written as $\Gamma(-{s\over2})\over\Gamma({s\over2})$ and gives \behavphase. We reached an important conclusion, that the phase \behavphase\ comes from the $\phi$ zero mode
integral.\foot{The role of the non-zero mode path integral is to cancel the imaginary part of $\delta$ (\ie\ the modulus of the reflection coefficient); in particular,
it presumably takes care of the real exponent in \sa.}

The next question that we would like to address is which region in the zero mode integral gives rise to the large phase shift \behavphase. To do that, we write \intzero\ as
\eqn\estint{\int_{-\infty}^\infty d\phi_0 A(\phi_0)e^{2ip_\phi\phi_0}~,}
where $p_\phi=-i(\beta+Q)=2Q\hat s$ is the radial momentum of the incoming particle \pphhii, and
\eqn\aalpha{
A(\phi_0)=\exp\left(-\alpha e^{-\phi_0/Q}\right)~, }
with
\eqn\hjh{\alpha=\lambda \int e^{b\phi_1}\cos R(x_L-x_R)~.}
For $\alpha <0$, $A(\phi_0)$
grows rapidly at negative $\phi_0$, and the integral diverges;
it is defined via analytic continuation from $\alpha > 0$. For $\alpha > 0$, we have the following situation. As $Q\to0$, $A(\phi_0)$ approaches a step function that vanishes for $\phi_0 < 0$, and thus for any value of $p_\phi$ the contribution to \estint\ comes from $\phi_0 > 0$. This agrees with the  picture coming from the semiclassical cigar geometry, according to which the radial direction lives on a half line. Indeed, it is natural to relate $\phi_0=0$ to the tip of the cigar.

For small but finite $Q$, the step function is smoothed out. For $\phi_0> Q$, $A(\phi_0)$ is still essentially a constant, and for large $p_\phi$ the oscillation of the phase in \estint\ leads to large cancellations. Hence, the region $\phi_0> Q$ does not contribute to the integral \estint. For negative $\phi_0$, $A(\phi_0)$ goes rapidly to zero, so the contribution from negative $\phi_0$ is limited to a finite range. One way to estimate this range is to compare $A(\phi_0)$ with the value of the integral that for large $p_\phi$ is given by  \sa; $\phi_0$ such that $A(\phi_0)$ is much smaller than \sa\ do not contribute much to the integral. This implies that the region $\phi_0<-Q \log(p_{\phi})$ does not contribute to the integral.

We conclude that the main contribution to the large phase shift comes from the region
\eqn\region{-Q \ln p_{\phi} < \phi_0 < Q~.}
This can be verified by noting that for large $p_{\phi}$, \estint\ admits a saddle point away from the real line at
\eqn\saddlept{\phi_0/ Q  \sim i \pi /2  -\ln p_{\phi}~,}
where `$\sim$' stands for an equality up to sub-leading correction at large $p_\phi$. Plugging \saddlept\ back into \estint\ reproduces \sa.

The region \region\ is small for small $Q$, but it grows in the negative $\phi_0$ direction as $p_\phi$ increases. This means that the scattering wave can penetrate into the classically forbidden region behind the tip of the cigar, by an amount that increases with the radial momentum $p_\phi$ -- the hard wall of the cigar geometry is replaced by a soft one at high energy.

What does the region \region\ mean in the cigar geometry? In the Sine-Liouville model, the dilaton behaves like $\Phi\simeq -Q\phi_0$. Comparing with the second line of  \sltwo, we see that the Sine-Liouville coordinate $\phi_0$ and the cigar coordinate $r$ are related via
\eqn\relation{Q \phi_0 = \ln\cosh r~.}
Taking into account the fact that the metric \sltwo\ is $ds^2=dr^2/Q^2+\cdots$, we see that  the upper bound in \region, $\phi_0\simeq Q$, corresponds in the cigar variables to $r\simeq Q$, which is a distance of order $l_s$ from the tip. The lower bound in \region\ corresponds to negative $\phi_0$, which does not exist in the classical cigar geometry.

We see that the full geometry corresponding to the coset model is well described by the cigar background \sltwo\ for distances larger than $l_s$ from the tip, and has a second asymptotic region $\phi_0<0$ which does not exist in the cigar description. The large phase shift \behavphase\ comes from this second region. The two regions are connected at a distance of order $l_s$ from the classical tip of the cigar.   This agrees with the fact that the wound tachyon on the cigar has support in a region of size $l_s$ from the tip \refs{\KutasovRR\GiveonKP-\GiveonICA}.

Finally, we note that the relation \relation\ maps the wound tachyon profile \lsinel\ to
\eqn\tacshape{e^{ -\phi_0/Q} \sim {1 \over \cosh^k(r)}~,}
which is precisely what one finds by solving the Klein-Gordon equation for the tachyon on the cigar\foot{In the notation of \GiveonICA, $\rho =\sqrt{\alpha' k} r$, $k=\alpha'/Q^2$.} \GiveonICA. In the bosonic case, this is valid for large $k$, while in the supersymmetric case the agreement is exact.

\newsec{Summary}
To summarize, the main points of this note are:
\item{(1)} The scattering phase shift in the full classical string theory on the cigar has a qualitatively different large radial momentum behavior from the one obtained in the gravity approximation. Even when the cigar is large and weakly curved (\ie\ for large $k$), for sufficiently large momenta the exact answer exhibits a growing scattering phase shift \behavphase, while the one obtained in the gravity approximation goes to zero.
\item{(2)} Even though the momenta in question are stringy, one can still interpret the large scattering phase shift in terms of potential scattering. This is the statement that the scattering phase is due to the dynamics of the zero mode of the worldsheet field associated with the radial direction, $\phi$, rather than to fluctuations of the non-zero modes.
\item{(3)} The large phase shift comes from a region behind the tip of the cigar, that does not exist in the classical geometry. In this region, the string sees a different background, corresponding to Sine-Liouville (or, in the supersymmetric case, $N=2$ Liouville). The two regions are connected at a distance of order $l_s$ from the classical tip of the cigar. 
\item{(4)} The FZZ duality between the cigar and Sine-Liouville manifests itself at large $k$ through the energy dependence of the physics. At low energies the cigar description is good, but at high energy the physics is better described by Sine-Liouville.

\noindent
Although the detailed results of this note involved a specific black hole background, we expect the basic picture to hold much more generally; in particular, it should apply to four dimensional Schwarzschild black holes. The fact that string theory in EBH backgrounds involves a non-zero condensate of the normalizable mode of the winding tachyon is believed to be general \KutasovRR. Moreover,  in  \refs{\GiveonKP,\GiveonICA,\GiveonHFA} it was argued that the presence of normalizable states near the Euclidean horizon is a general property of EBH spacetimes. It is natural to expect that the FZZ duality and its implications for the geometry seen by perturbative strings at various energies discussed in this note hold in general as well.

\bigskip
\noindent{\bf Acknowledgements:}
We thank R. Ben-Israel, S. Elitzur, J. Lin, E. Martinec, E. Rabinovici and J. Teschner for discussions. This work was supported in part by the BSF -- American-Israel Bi-National Science Foundation.
The work of AG and NI is supported in part by the I-CORE Program of the Planning and Budgeting Committee and the Israel Science Foundation (Center No. 1937/12), and by a center of excellence supported by the Israel Science Foundation (grant number 1989/14). DK is supported in part by DOE grant DE FG02-13ER41958. DK thanks Tel Aviv University and the Hebrew University for hospitality during part of this work.

\listrefs

\end